\documentclass[final,5p,times,twocolumn]{elsarticle}

\usepackage{amssymb}

\usepackage{lineno}
\usepackage{graphicx,dblfloatfix}
\usepackage{multirow}
\usepackage{booktabs}
\usepackage{subcaption}
\usepackage{longtable}
\usepackage{amsmath}
\usepackage{fancyhdr}
\usepackage{multirow}
\usepackage{color}
\usepackage{geometry}
\usepackage{rotating}
\usepackage{lscape}
\usepackage{comment}
\usepackage{fancyvrb}
\usepackage{listings}
\usepackage{afterpage}
\usepackage{float}

\usepackage{placeins}

\journal{SoftwareX}

\usepackage{hyperref}

\begin{document}

\begin{frontmatter}



\title{MODULO: A software for Multiscale Proper Orthogonal Decomposition of data}


\author{Davide Ninni\fnref{nt1}}
%

\author{Miguel A. Mendez\corref{cor}}
\ead{mendez@vki.ac.be}

\cortext[cor]{Corresponding author}

\fntext[nt1]{Currently at Politecnico di Bari, Dipartimento di Meccanica, Matematica \& Management, 70125 Bari, Italy.}

\address{von Karman Institute for Fluid Dynamics, EA Department, 1640 Sint-Genesius-Rode, Belgium}

\begin{abstract}
In the era of the Big Data revolution, methods for the automatic discovery of regularities in large datasets are becoming essential tools in applied sciences. This article presents an open software package, named MODULO (MODal mULtiscale pOd), to perform the Multiscale Proper Orthogonal Decomposition (mPOD) of numerical and experimental data. This novel decomposition combines Multi-resolution Analysis (MRA) and standard Proper Orthogonal Decomposition (POD) to allow for the optimal compromise between decomposition convergence and spectral purity of its modes. The software is equipped with a Graphical User Interface (GUI) and enriched by numerous examples and video tutorials (see Youtube channel \href{https://www.youtube.com/watch?v=ED3x00H4yN4&list=PLEJZLD0-4PeKW6Ze984q08bNz28GTntkR}{\textcolor{blue}{MODULO mPOD}}). \color{blue} The \textsc{Matlab} source codes and an executable for Windows users can be downloaded at \color{blue}\url{https://github.com/mendezVKI/MODULO/releases}; a collections of exercises in Matlab and Python are provided in \color{blue}\url{https://github.com/mendezVKI/MODULO}. \color{black}
\end{abstract}


\end{frontmatter}



\section{Motivation and significance}\label{Sec1}

Data driven modal analysis aims at decomposing a dataset as a linear combination of elementary contributions called \emph{modes}. This provides the fundamental framework for many areas of applied mathematics and related applications, including pattern recognition, machine learning, data compression, filtering and model order reduction \cite{TechnologyStrang2019,Brunton2019}. \color{blue} Linear decomposition consists in projecting the dataset onto a suitable space\color{black}, spanned by a basis that is, hopefully, better capable of capturing the essential features of the data.

A \emph{mode} is computed by projecting the data onto a certain element of a basis. In most of the engineering applications, the data results from the discretization, or sampling of a real quantity (e.g. grayscale entries in images, pressure fields in a fluid flow simulation or stress fields in solid mechanics) over a spatial discretization $\mathbf{x}_i$ and a temporal discretization $t_k$. Each of the modes produced by a decomposition has its own spatial and temporal structure.

In pattern recognition, one aims at linking modes to specific patterns of interest \cite{Bishop2006,Bouwmans2017}.
In filtering or data compression, one aims at removing modes that describe unwanted features or modes that do not significantly contribute to the data \cite{Lu2011,Davydenko2014}. In machine learning, this is often a fundamental pre-processing step for many supervised or unsupervised problems \cite{Alpaydin2014}. In model order reduction, one aims at projecting partial differential equations onto the space spanned by a few of the leading modes \cite{Keiper2018,Quarteroni2016}, thus significantly reducing the computational complexity of a problem and eventually enabling system identification methods and control \cite{Reynders2012,Noack2011}.

Contrary to classical tools such as the Discrete Fourier Transform (DFT) or Discrete Wavelet Transform (DWT), data-driven decompositions tailor their bases to the data investigated. The Proper Orthogonal Decomposition (POD, \cite{Holmes1996}), also known as Principal Component Analysis (PCA, \cite{Jackson2003}) or Karhunen-Lo\'{e}ve transform \cite{Ortho} depending on the field, is the most classic example. This decomposition is often implemented using the Singular Value Decomposition (SVD) and constructs its bases such that the error produced by truncating the decomposition to any $r<R$ is minimized.

The convergence optimality comes at the cost of setting no constraints on the frequency content of the structures constituting its modes. In many applications, however, it is of interest to have harmonic decompositions to facilitate physical interpretability. In fluid mechanics, this need has motivated the development of an alternative decomposition known as Dynamic Mode Decomposition (DMD, \cite{Kutz2014}), which extends the Fourier decomposition to the data-driven paradigm: frequencies are defined from the dataset and not imposed a priori.

There exist nevertheless cases for which both convergence optimality and spectral purity yields poor feature detection capabilities \cite{Mendez2019}. The lack of frequency constraints in the POD occasionally results in modes that capture phenomena occurring at very different scales. On the other hand, the constraint of purely harmonic modes prevents time-frequency localization and yields convergence problems for datasets that are not strictly periodic.

The Multiscale Proper Orthogonal Decomposition proposed in \cite{Mendez2019} allows for overcoming the limitations of the two with a hybrid method that combines their advantages. This decomposition has already been successfully used in various experimental \cite{MendezJournal2,Mendez20192,DANTEC_VKI,CLAU} and numerical works \cite{Amor2019,Mendez2019}. The software package \textit{MODal mULtiscale pOd} (MODULO) described in this work is a Matlab-based software developed at the von Karman Institute for Fluid Dynamics to perform mPOD, POD and DFT on numerical and experimental data. Equipped with a Graphical User Interface (GUI), an executable, and a set of exercises and video tutorials, the use of this software does not require direct interaction with the source code and can thus facilitate applied scientists that are unwilling to enter into technicalities linked to programming. In what follows, we refer the reader to each of the eight video tutorials in MODULO's \href{https://www.youtube.com/watch?v=ED3x00H4yN4&list=PLEJZLD0-4PeKW6Ze984q08bNz28GTntkR}{\textcolor{blue}{ youtube channel}} for a more detailed review of the software features.






\section{Software Description}\label{SEC2}

\subsection{Theoretical Background}\label{SEC21}
A brief description of the theoretical background of the code is here \color{blue}presented\color{black}; for more details, the reader is referred to \cite{Mendez2019} and to the first three video tutorials in the \href{https://www.youtube.com/watch?v=ED3x00H4yN4&list=PLEJZLD0-4PeKW6Ze984q08bNz28GTntkR}{\textcolor{blue}{youtube channel}}. As described in the
\href{https://www.youtube.com/watch?v=ED3x00H4yN4}{\textcolor{blue}{first video}}, all the modal decompositions implemented are matrix factorizations of the form

\begin{equation}
{D}(\mathbf{x_i},t_k)
=\Phi(\mathbf{x_i})\,\Sigma\,\,\Psi(t_k)^T
\label{UNO}\end{equation} where ${D}$ is the data matrix to decompose, here assumed to be function of space ($\mathbf{x}_i$) and time ($t_k$). \color{blue}Regardless of the dimensionality (e.g. 2D or 3D, scalar or vector) of the data, we here assume that every temporal realization (snapshot) of the data is flattened into a vector and stored as a column of $D$. Every column of is thus a function of $\mathbf{x}_i$ and every row is a function of $t_k$. \color{black} This matrix is thus of size ${n_s \times n_t}$, with $n_s$ the number of spatial points and $n_t$ is the number of temporal realizations (snapshot), and has rank $R=rank(D)\leq N= min(n_s,n_t)$.

The matrices $\Phi\in{\mathbb{C}^{n_s\times R}}$ and $\Psi\in{\mathbb{C}^{n_t\times R}}$ collect the spatial and the temporal structures (bases) and $\Sigma=diag[\sigma_1,\sigma_2,\dots,\sigma_{R}]\in{\mathbb{R}^{R\times R}}$ is the diagonal matrix collecting their importance (amplitude). More generally, if other independent variables are considered instead of space and time, $\Phi$ and $\Psi$ contain a basis for the columns and the rows of $D$ respectively.

\color{blue}
The bold notation used ($\mathbf{x}_i$ or $\mathbf{i}$) denotes linear matrix indices. In the current version of the code, it is assumed that the data has uniform sampling both in space and in time, although the latter constraint can be \color{blue}relaxed \color{black}for the POD. Consider, for example, a 2D velocity field from planar Particle Image Velocimetry (PIV) sampled over a grid $\mathbf{x_i}=[x_i,y_j]$ containing $128\times 128$ points. Flattening the entire velocity field into the columns of $D$, the number of spatial points if $n_s=2\times 128\times 128= 16384$, having concatenated both vector components into a single snapshot.

\color{black}

The matrix ${D}$ is constructed by assigning every $k^{th}$ snapshot $d_k[\mathbf{i}]$ to a column of $D$. Here $k$ is the index over the time discretization $t_k=(k-1)\Delta t$, sampled at a frequency $f_s=1/\Delta t$ with $k=[1,n_t]$. As the spatial structures $\phi_r[\mathbf{i}]$ are columns of $\Phi$, and the temporal structures $\psi_r[k]$ are columns of $\Psi$, eq.\eqref{UNO} can be written as a dyadic expansion:

\begin{equation}
D[\mathbf{i},k]=\sum_{r=1}^{N}\,\sigma_r\phi_r[\mathbf{i}]\,\psi_r^T[k]\,.
\label{summation}
\end{equation} If the summation is truncated at $r<N$, an approximation of the original data is obtained. Since $\Sigma$ is a diagonal matrix, and both the spatial and temporal structures have unitary norm $||\phi_r||=||\psi_r||=1\,\forall r\in[1,R]$, it is easy to see that the decomposition can be completed once either $\Phi$ or $\Psi$ are given.

The decompositions implemented in MODULO are the POD, the DFT, and the mPOD, computing first the temporal structures $\Psi$. All these decompositions have an orthonormal temporal basis ($\Psi^{-1}=\Psi^\dag$, with $^\dag$ denoting Hermitian transpose). Hence, the spatial structures can be easily computed as

\begin{equation}
\label{eq3}
\Phi=D\,\overline{\Psi}\,\Sigma^{-1}\,,
\end{equation} where the calculation of the diagonal matrix $\Sigma$ is done from the normalization of the spatial structures, that is $\sigma_r=||D\psi_r||$.

The theoretical background for DFT and POD is provided in the \href{https://www.youtube.com/watch?v=8fhupzhAR_M}{\textcolor{blue}{second video tutorial}} while the \href{https://www.youtube.com/watch?v=eG17tRQxqhk}{\textcolor{blue}{third video tutorial}} is dedicated to the mPOD. \color{blue}The computation of the temporal structures for these three decompositions proceeds as follows.
\color{black}
\paragraph{-DFT} The temporal basis is the well known Fourier matrix, which can be written as $\Psi_{\mathcal{F}}(i,j)=w^{(i-1)\,(j-1)}$, with $w=e^{2\,\pi\,i/n_t}$ with $i=\sqrt{-1}$. \color{blue} The construction of this matrix is independent from the data, hence the DFT is not \emph{data-driven}. In practice, the multiplication by the Fourier Matrix is carried out using the FFT algorithm.\color{black}

\paragraph{-POD} The temporal basis is computed from the eigenvalue decomposition of $K=D^TD$:

\begin{equation}
\label{eq4}
    \Psi_{\mathcal P} : \rightarrow K=\Psi_P\,\Lambda\,\Psi^T_P.
\end{equation} This matrix is known as temporal correlation matrix in the fluid mechanics community and the approach implemented is known as Sirovinch's snapshot method \cite{Siro3}.

\paragraph{-mPOD} The temporal basis is computed via a combination of Multi-resolution analysis and eigenvalue decomposition. The fundamental idea of the mPOD is to perform POD at different scales, each retaining non-overlapping portions of the frequency spectra. For example, assuming that the dataset is sampled at $f_s=1000\,Hz$, one might decide to separate phenomena occurring in the range $[0-100]\,\,Hz$, $[100-300]\,\,Hz$ and $[300-500]\,\,Hz$ and perform a POD on each of these independently. As described in \cite{Mendez2019}, this could be done by using a filter bank, defined by a frequency splitting vector $F_V=[100, 300]\,\,Hz$, to break the dataset into three contributions, and perform the POD in each of these separately. To reduce the computational cost of this operation, the mPOD performs the MRA on the temporal correlation matrix $K={D}^TD$. Given a set of suitable transfer functions $\{\mathcal{H}_m\}^{M}_{m=1}$, with $M$ the number of scales to identify, the mPOD breaks the correlation matrix as:

\begin{equation}
    \label{K_MPOD}
    K=\sum^{M}_{m=1}K_m=\sum^{M}_{m=1}\,\Psi_{\mathcal{F}}\,\Bigl [\widehat{K}\odot \mathcal{H}_m\Bigr] \Psi_{\mathcal{F}}\,
\end{equation}

where $\widehat{K}=\overline{\Psi}_{\mathcal{F}}\,K\,\overline{\Psi}_{\mathcal{F}}$ is the 2D Fourier transform of the correlation matrix and $\odot$ is the \color{blue}shur 
\color{black}   product, that is the entry by entry multiplication.

The filtering operation is designed to preserve the key properties of $K$: each contribution $K_m$ is symmetric and positive definite and thus equipped with a set of orthonormal eigenvectors (POD modes) and non-negative real eigenvalues:

\begin{equation}
 K_m=\Psi_{\mathcal{P}m}\Lambda_{m}\Psi_{\mathcal{P}m}^T=\sum^{n_m}_{j=1}\lambda_{m}\psi_{\mathcal{P}\,m\,j}\psi_{\mathcal{P}\,m\,j}^T\,,
\end{equation} where $n_m$ is the number of non zero eigenvalues at each scale.

These spectral constraints impose that a mode (eigenvector) having frequency content in one scale has no frequency content in the others. Therefore, it is possible to show that the eigenvectors of all the scale are orthogonal complements that span the entire $\mathbb{R}^{n_t}$ space, that is $\sum^{M}_{m=1}{n_m}\approx n_t$.

The mPOD basis is then constructed by collecting the POD bases of all scales, sorted by amplitude:

\begin{equation}
    \Psi_{\mathcal{M}}=[\Psi_1,\Psi_2\,\dots\Psi_M]P_{\Sigma}
\end{equation}\, with $P_{\Sigma}$ a permutation matrix to rank the structures in decreasing order of energy contribution.
This decomposition generalizes POD and DFT: for $M\rightarrow 1$, the mPOD is a standard POD. At the limit $M\rightarrow n_t$, the mPOD is a DFT.

\subsection{Software Architecture}

\begin{figure}[h]
\center
\includegraphics[width=5cm]{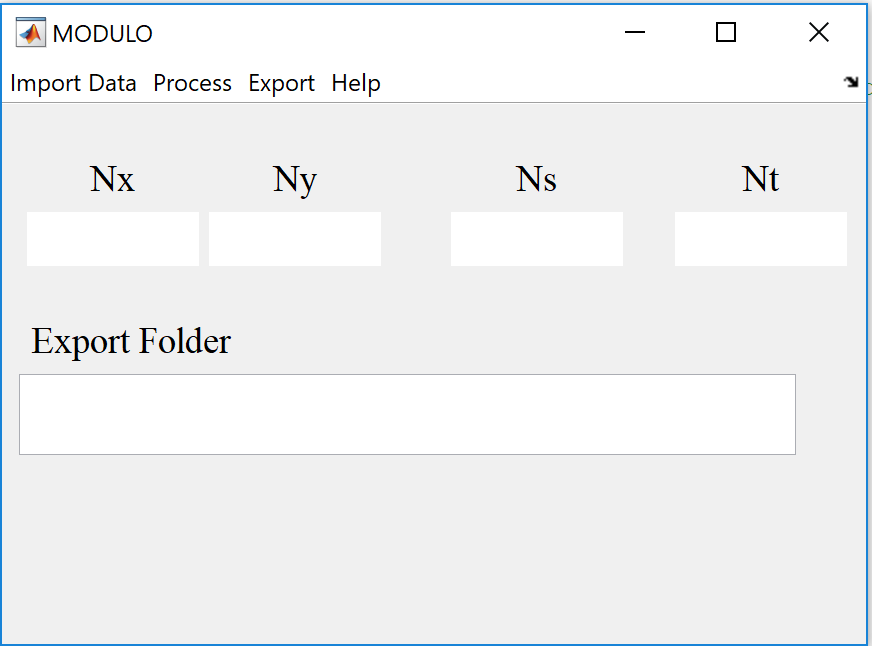}
\caption{Main menu of MODULO. Here the user sets the exporting folder and proceeds through the decomposition from the upper toolbar.}
\label{fig:main_menu}
\end{figure}

MODULO has a minimal Graphical User Interface (GUI), shown in figure \ref{fig:main_menu}, that opens after launching the executable. The decomposition process can be followed from the toolbar in the upper part of the main menu. An overview of MODULO's GUI is given in the \href{https://www.youtube.com/watch?v=s1ER4erdpsc}{\textcolor{blue}{fourth video tutorial}}.

In this first version of the software, the data is assumed to be 1D or 2D: the first raw of non-editable tabs shows the number of points along the x-axis (\verb+Nx+), the y-axis (\verb+Ny+), the total number of spatial points (\verb Ns) and the number of time steps (\verb+Nt+). In a 1D test case, it is assumed that \verb+Ns=Nx (Ny=0)+; in a 2D scalar test case one has \verb+Ns=Nx+$\times$\verb+Ny+ while a 2D vectorial test case yields \verb+Ns=2+$\times$\verb+Nx+$\times$\verb+Ny+.

The dataset can be imported from the menu \verb+Import Data+, which allows for two options: \textit{embedded mesh} or \textit{separated mesh}. The first option should be used if the mesh grid is stored in each of the data files; the second option should be used if the mesh is stored in another file.

\begin{figure}[h]
\center
\includegraphics[width=7cm]{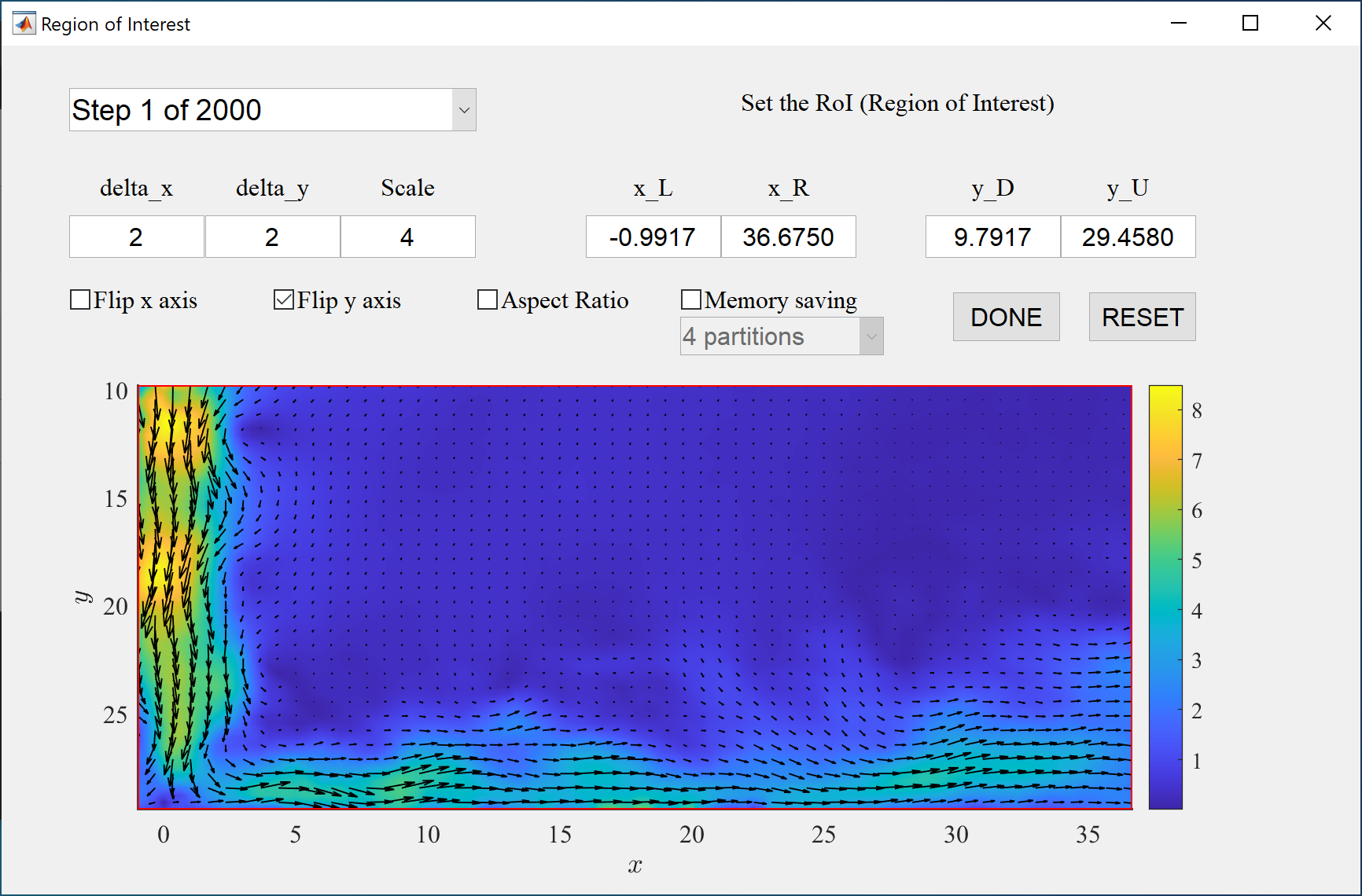}
\caption{Region of Interest window, to visualize the imported snapshots, select the portion of the domain that will be analized and several plotting parameters.}
\label{fig:roi}
\end{figure}

Before importing the data, the user can mean-center all the snapshots, i.e., remove the average column (time average if the row domain is linked to time) from the snapshot matrix $D$. This option can be useful for plotting purposes in the DFT and is generally recommended for POD and mPOD of statistically stationary datasets. Once the files are loaded, the \textit{Region of Interest} (RoI) GUI shown in figure \ref{fig:roi} opens.

This GUI allows for setting the portion of the spatial domain that will be used in the decomposition. By default, MODULO considers the entire domain, but the user can introduce the ranges along the horizontal (\verb+x_L, x_R+) and the vertical (\verb+x_L, x_R+) axes. The upper left menu pop-up menu allows for selecting and preview any of the imported snapshots. In case of vector datasets, the parameters on the left allow for adjusting the quiver plot in terms of spacing (\verb+delta_x, delta_y+) and arrow length (\verb+Scale+) while the bottom tick boxes can be used to flip the axes or adjust the axis aspect ratio to $1:1$. The settings used at the step will be used in the exporting of the spatial structures of the decomposition.

Once these parameters are set, the user can either use the \verb+RESET+ button, to restore default values or the button \verb+DONE+, to proceed with the importing of the data and the preparation of the matrix $D$. The \emph{Memory saving} option is described in the section \ref{sec:memory}.

\setcounter{figure}{3} \renewcommand{\thefigure}{\arabic{figure}} 
\begin{figure*}[!hb]
\center
\includegraphics[width=12cm]{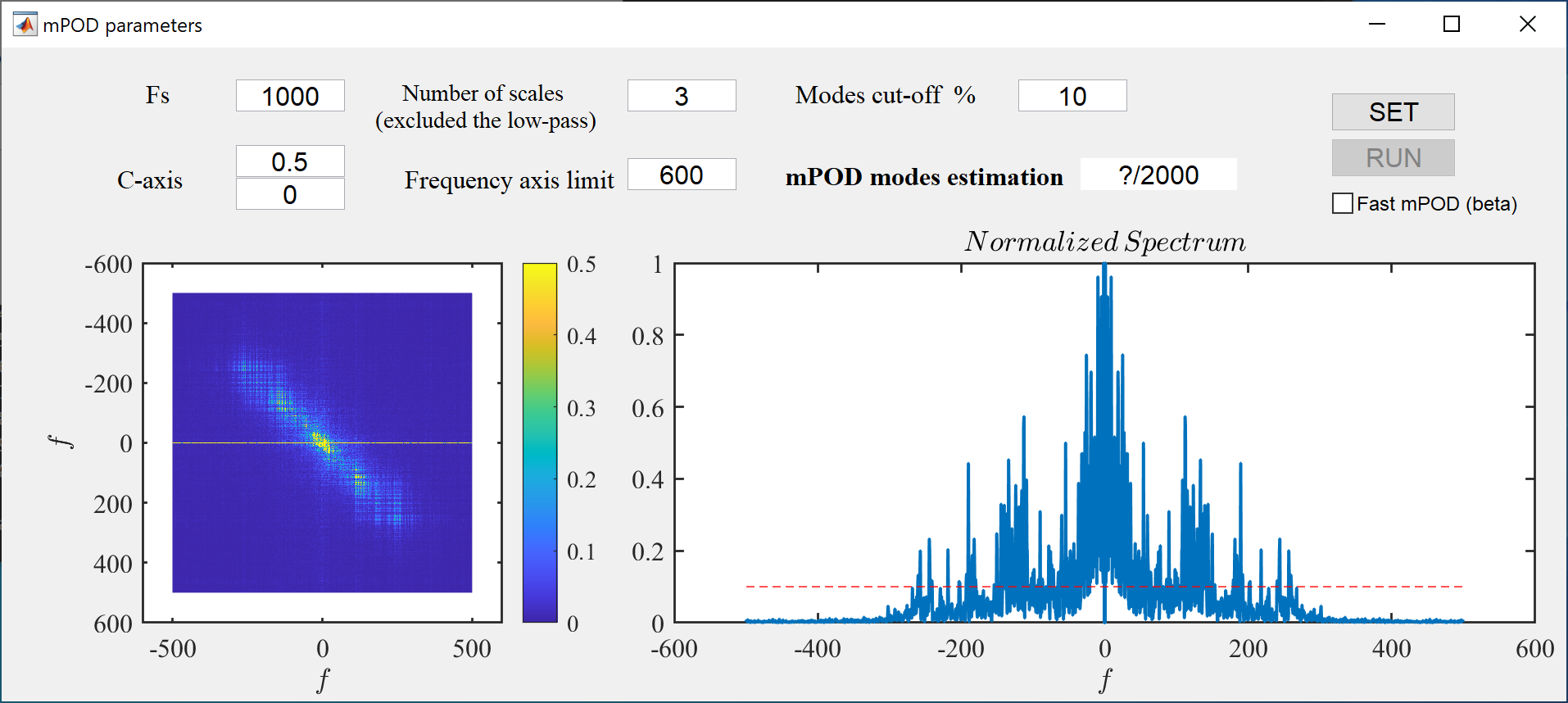}
\caption{GUI for introducing the mPOD parameters. During the setting definition, the user can monitor a contour of power spectral density matrix and its diagonal.}
\label{fig:mpod_par}
\end{figure*}

\subsection{Process}

From the \verb+Process+ menu, the user can select which of the three decompositions described in section \ref{SEC2}, is to be performed. By choosing POD or DFT algorithms, the user is asked to introduce the sampling frequency $f_s$ and the extreme of the range of the modes to be exported.
For all the decompositions, this indexing assumes that the modes are always exported in decreasing order of amplitude, even if DFT and mPOD are not energy-based. Nevertheless, for DFT and mPOD, the need for ordering the modes requires the calculation of the complete basis regardless of the number of exported modes, while for the POD only the modes to be exported are computed.

MODULO is a dimensionless software, hence the units in the sampling frequency need not to be specified: digital frequency bins are computed as $\Delta f=f_s/n_t$ and the frequency domain is $f_n\in[-f_s/2, f_s/2]$. Once these parameters are introduced, the decomposition begins in the case of DFT and POD, and a wait bar indicates the progress of the calculation. The
\href{https://www.youtube.com/watch?v=MUrkOmI2aaw}{\textcolor{blue}{fifth}} and
\href{https://www.youtube.com/watch?v=TVJX6JNQASU}{\textcolor{blue}{sixth video tutorials}} are dedicated to the POD and the DFT respectively.

In the case of mPOD, additional parameters should be introduced, and a dedicated GUI appears to support the user in this process. This GUI is described in the \href{https://www.youtube.com/watch?v=srcU_jfalXY}{\textcolor{blue}{seventh video tutorial}} and is shown in figure \ref{fig:mpod_par}.

The window shows a contour of the value of the cross-spectral density matrix on the left and its diagonal on the right. On the top-left editable box, the user introduces the sampling frequency \verb+Fs+ and the number of scales \emph{excluding the largest}. Assuming, for example, that one is interested in four scales, say $[0-100]Hz$, $[100-200]Hz$, $[200-300]Hz$, and $[300-Fs/2] Hz$, the number to introduce here is $3$. The reason for excluding the first (largest) scale from the counting is that this scale is always kept by default in MODULO while the other scales can be removed if the software is used as a low-pass filter.



The parameters \verb+C-axis+ and \verb+Frequency axis limit+ can be modified for plotting purposes; the first changes the upper limit in the color axis of the contour plot, the second sets the frequency limits in the axes of both figures.


The last input parameter in the figure is the \verb+Modes cut-off+, which controls the number of modes computed as indicated in the non-editable text box below. A red line also indicates this percentage in the normalized spectrum (right plot). This number controls the number of modes, in each scale, that will be considered in the final mPOD basis: if this is set to, e.g., $10\%$, then only the modes that have at least $10\%$ of the leading POD mode energy in each scale will be taken. Observe that this estimation is made \emph{before} computing the decomposition based on the transfer function of the filters that will isolate the scales, and is thus only an estimation. This estimation is available only after the settings of the filters in each scale are introduced, from the \verb+SET+ button. The first parameter is the frequency splitting vector: its entries collect the cut-off frequencies on each scale. In the previous example this is $F_V=[100,200,300]Hz$ (units are customary).
The second input is a vector containing the \textit{length of the filter kernels} that will be used to compute each scale. These are FIR filters with a Hamming window. Finally, the last input allows for introducing the \textit{keep vector}. This indicates the intermediate scales that will be kept and allows for using the mPOD also as a filter. In the previous example, if the keep vector is set to $[1,1,0]$, then the highest scale $[300-Fs/2] Hz$ is removed from the decomposition. If the keep vector is set to $[0,0,0]$ then the mPOD will be equivalent to performing the POD of the dataset obtained by low-pass filtering to keep only the portion $[0-100]Hz$.


Once the \textit{frequency splitting vector}, the \textit{length of the kernels} and the \textit{keep vector} are set, the \verb+RUN+ button to start the computation is enabled. As for the DFT, the user inputs the range of modes to be exported only at the end of the decomposition.

\subsubsection{Memory Saving}\label{sec:memory}
The largest matrix in every decompositions is the snapshot matrix $D\in\mathbb{R}^{n_s\times n_t}$. Depending on the computational resources available, this matrix can be prohibitively large, and the cost of matrix products such as the correlation matrix in \eqref{eq4} or the projection in \eqref{eq3} might exceed the available RAM. To cope with this limitation, MODULO offers a `Memory Saving' option, from the RoI GUI shown in Figure \ref{fig:roi}. This option is described in the  \href{https://www.youtube.com/watch?v=s1ER4erdpsc}{\textcolor{blue}{eighth video tutorial}}.

\begin{figure}[htb]
\centering
	\includegraphics[width=8cm]{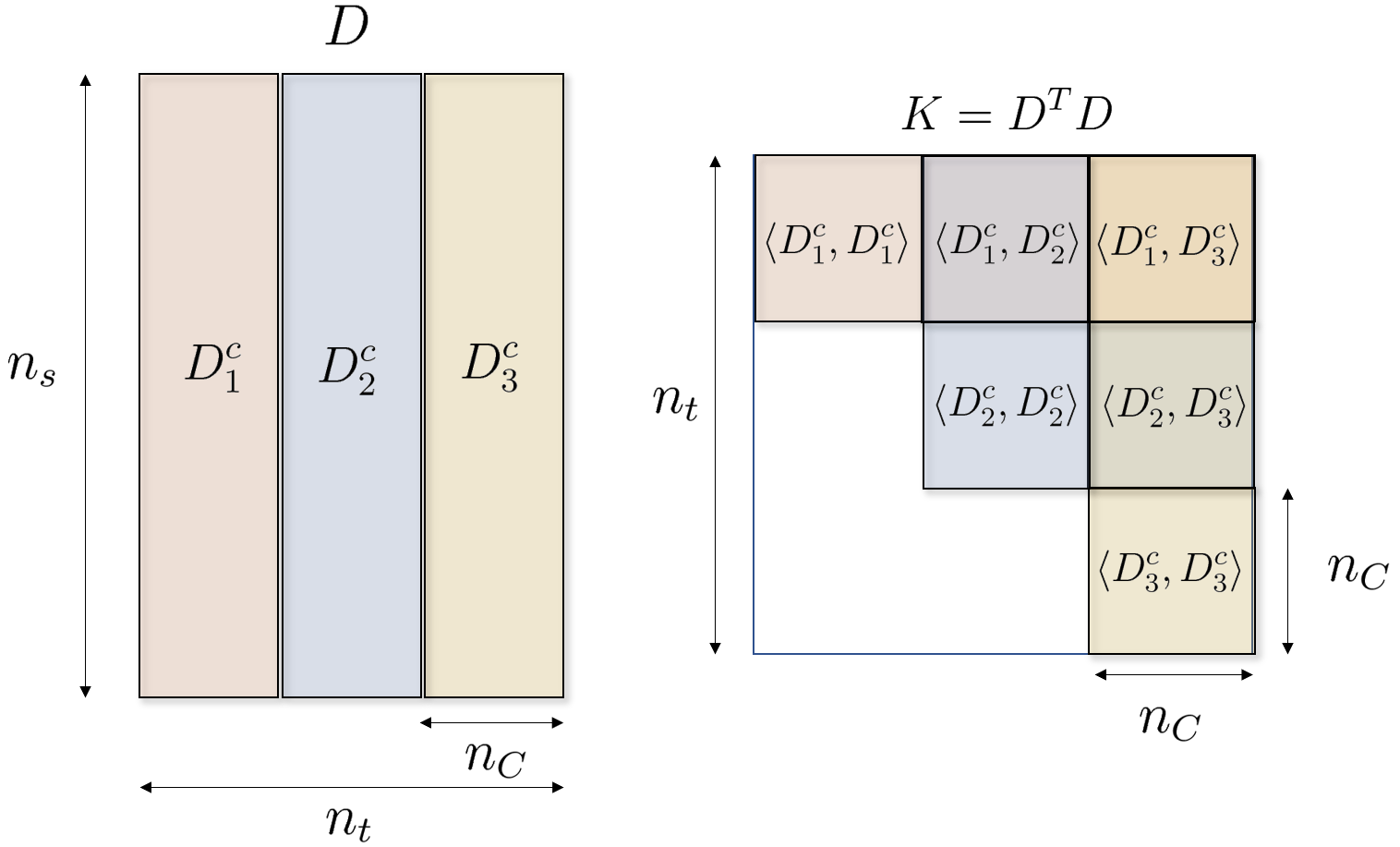}
\caption{Calculation of the temporal correlation matrix in \eqref{eq4} using three partitions to limit memory requirements. Only half of the blocks needed are computed; the remaining portion being symmetric.}
\label{fig:CovK}
\end{figure}

When this option is active, all decompositions are performed without loading the snapshot matrix $D$ in memory, but only a few partitions of it at a time. Therefore, the calculation of the time correlation matrix $K=D^TD$, needed for POD and mPOD, is computed from $n_P$ `column-wise' partitions of $D$. Each partition $D^c_i$ is of size $n_s\times n_C$ with $n_C=n_t/n_P$. These partitions are saved as temporary \textit{.mat} files and loaded one at a time during the calculation of the correlation, which is performed in blocks. This allows for limiting the number of stored entries to $n_s \times n_C\times 2$ entries, although at the cost of increasing the number of reading/writing operations. The calculation steps for this matrix using three blocks is illustrated in Figure \ref{fig:CovK}. To limit the cost of the memory saving feature, MODULO takes advantage of the symmetry of the correlation, computing only the upper triangular part of it while the remaining portion is mirrored.

\begin{figure}[htb]
\centering
	\includegraphics[width=8cm]{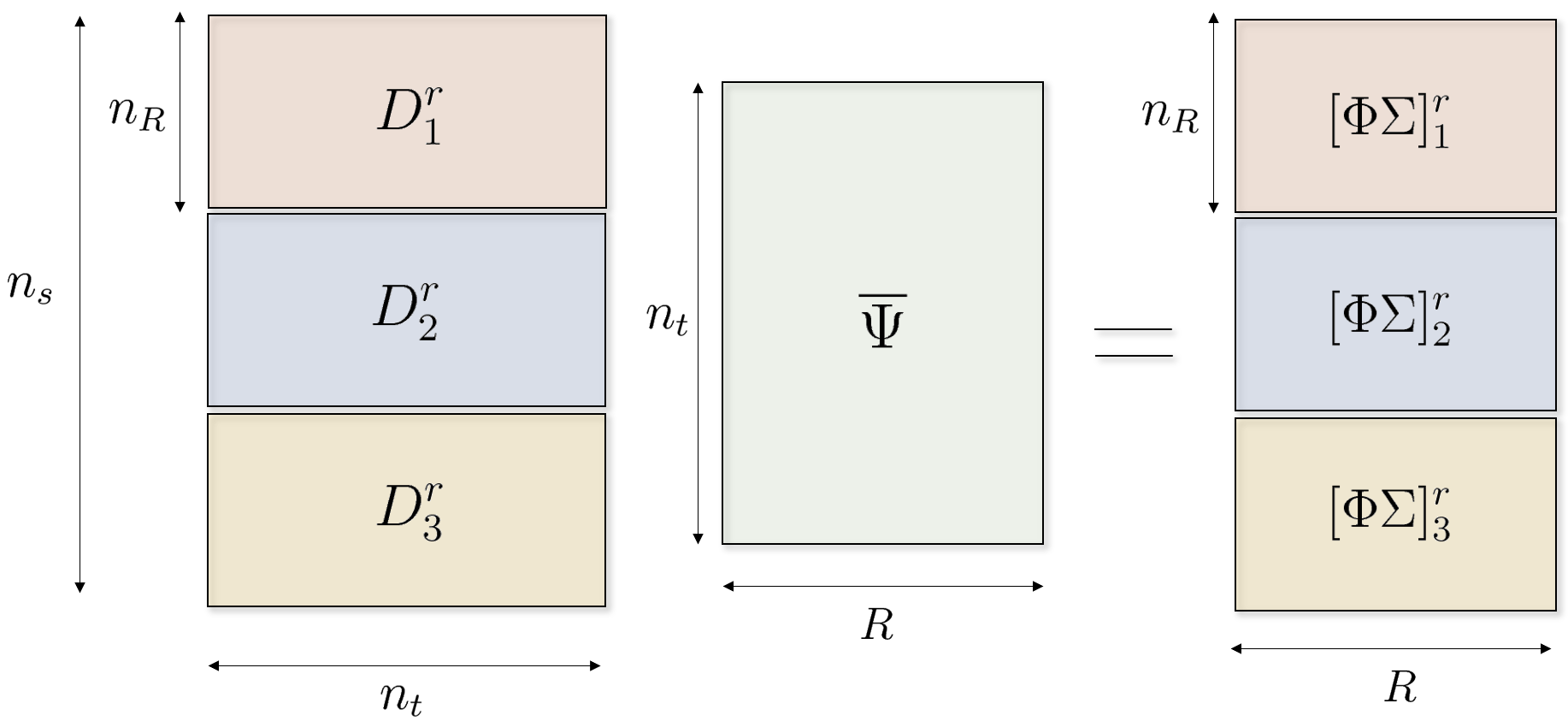}
\caption{Calculation of the final projection in \eqref{eq4} using three partitions to limit memory requirements. The final 'row-wise' blocks of $\Phi \Sigma$ must be regrouped into 'column-wise' blocks for the normalization step.}
\label{fig:PhiSigma}
\end{figure}

A similar approach is pursued in computing \eqref{eq3}, namely the last projection step of every decomposition. In this case, $D$ is split into $n_P$ `row-wise' partitions $D_r$, of size $n_R\times n_t$, and the projection is carried out independently in each portions. The calculation steps for this matrix using three blocks is illustrated in Figure \ref{fig:PhiSigma}. 
The resulting projection is then assembled back to column-wise partition to compute the amplitude of each mode via column normalization of the matrix $\Phi \Sigma$. 

In all the analyzed exercises, the increased computational cost of the memory saving option is of the order of a minute, depending on the decomposition and the size of each snapshot. Table \ref{tab:Comp} collects the computational time (in seconds) required to perform POD, DFT and mPOD without memory saving ($n_p=1$) and with memory saving with $n_p=4$ and $n_p=12$. The calculations performed on a laptop with Intel(R) Core(TM) i7-7700HQ CPU with $2.8 GHz$ and $16$ GB RAM.

\begin{table}[]
\center
\begin{tabular}{lcccc}
                                         & Time Steps & $n_p=1$ & $n_p=4$ & $n_p=12$ \\ \hline
\multicolumn{1}{c}{\multirow{3}{*}{POD}} & 500        & 0.56       & 12.38  & 17.22          \\
\multicolumn{1}{c}{}                     & 1000       & 1.20      & 23.23    & 32.22          \\
\multicolumn{1}{c}{}                     & 2000       & 4.97       & 50.87   & 68.19         \\ \hline
\multirow{3}{*}{DFT}                     & 500        & 3.37      & 29.49    & 37.12         \\
                                         & 1000       & 6.5      & 57.19    & 73.14         \\
                                         & 2000       & 14.39       & 110.96   & 149.78          \\ \hline
\multirow{3}{*}{mPOD}                    & 500        & 1.89      & 5.01      & 8.41          \\
                                         & 1000       & 7.80       & 13.39   & 20.46          \\
                                         & 2000       & 52.64       & 65.18     & 77.86          \\ \hline
\end{tabular}
\caption{Computational time (in seconds) of POD,DFT, and mPOD with no memory saving option ($n_p=1$) and with memory saving with different partitions. The test case considered is the one from Exercise 4 in MODULO's Github repository, with each snapshot consisting of a 2D velocity field on a $n_s=13680$ grid. The mPOD is performed with four scales. The time for preparing the dataset matrix is not included; for the mPOD, the timing also excludes the preparation of $K$.}
\label{tab:Comp}
\end{table}

The timing is given in output by MODULO. Hence the timing to prepare the dataset is excluded (since this step is done before selecting the decomposition), while for the mPOD this timing also excludes the preparation of the correlation matrix (since this is done before the mPOD setting GUI opens). The scope of Table \ref{tab:Comp} is thus not that of comparing the decomposition time but comparing how the partitioning influences the timing. For the mPOD, four scales are chosen with kernel widths of 100 and all scales being kept.

The most expensive operation appears to be the final projection with normalization and sorting steps: since the DFT always requires computing $n_t$ modes, this decomposition is more sensitive to the increase of the snapshots. On the other hand, as the POD requires no normalization, the computational costs are much reduced. In general, the computational time increases with the number of partitions due to the increased time spent in reading/writing operations. The price to pay to maintain a limited memory usage appears nevertheless acceptable. Finally, it is worth observing that in case the memory saving is not selected, but the computational resources are not sufficient for the calculation, a warning dialog appears. In this case, the user is strongly advised to either hit ``Activate Memory Saving'' (in which case the memory saving will be activated with the default number of partitions), either close the warning box and select one of the proposed partitions.

\subsection{Export}\label{sec:export}

Through the menu \verb+Export+, the exporting folder can be chosen. In this folder (which is created if not already available), the modes are saved as .png and .xlsx files. In particular, the excel files are always saved (for the mesh, for the sigmas, for the spatial structures and for the temporal structures) while the pictures are saved only if the tick (checkbox) in the menu \verb+Export+ is signed. In the case of DFT, the temporal structures are not saved since these are sinusoidal with fixed frequencies ($\Psi_{\mathcal{F}}$ is known a priori as recalled in section \ref{Sec1}). More information on the exported data is discussed in the video tutorials of each decomposition.

\section{Illustrative Examples}\label{sec:examples}

\begin{figure}[htb]
\centering
	\includegraphics[width=7cm]{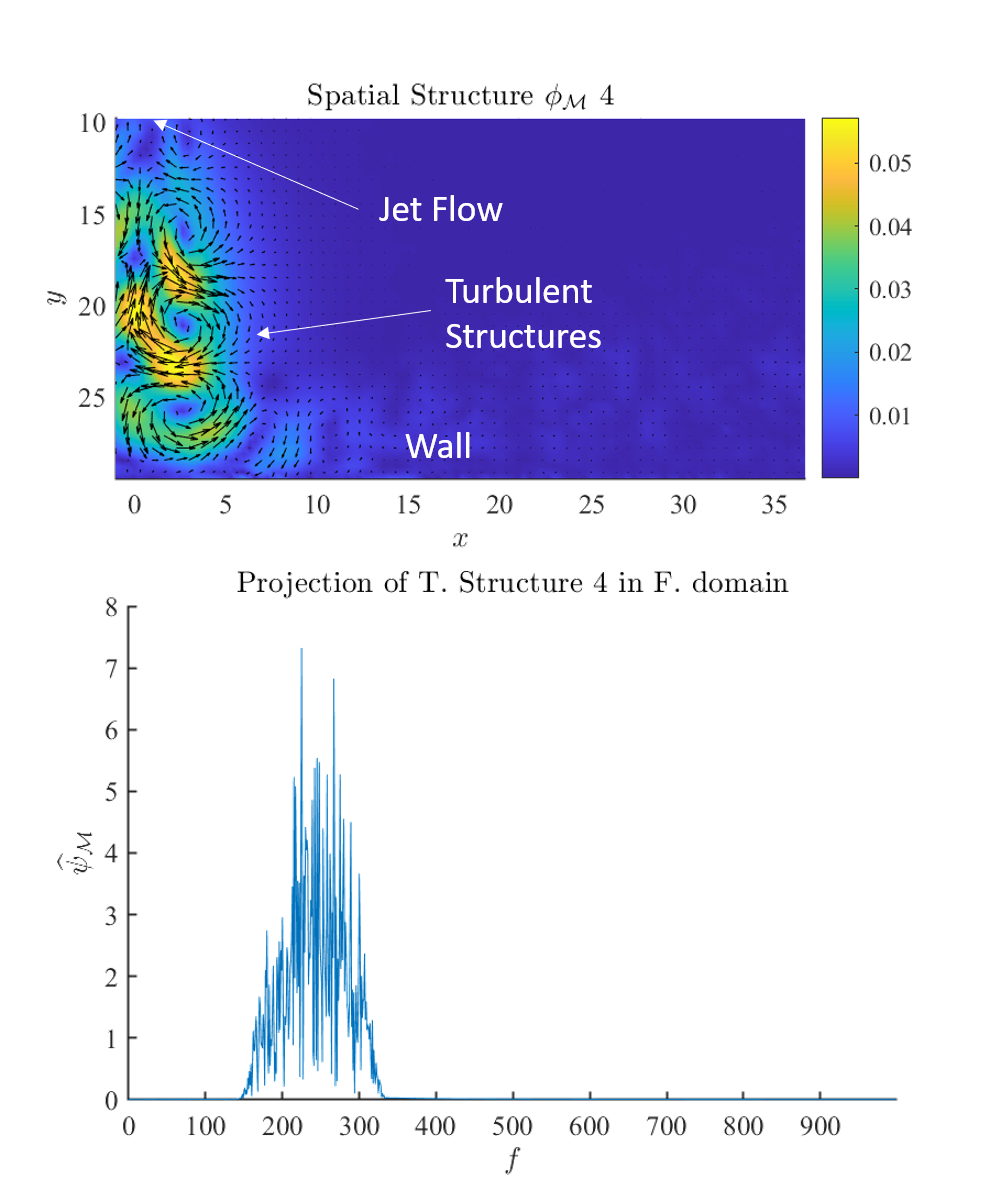}
\caption{Spatial structure and frequency content of the fourth mPOD mode in the TR-PIV velocity field of exercise four, consisting of a planar gas jet flow impinging on a flat wall.}
\label{fig:modes_jet}
\end{figure}

The code repository currently includes five exercises that allow for testing all the features of MODULO and, at the same time, explore the limitations and strengths of each decomposition. These exercises are also solved using various commented \textit{Matlab files} (sorted from `A to `D') in order to let the user follow the decomposition procedures using the source codes. The first exercise presents the analysis of a 1D scalar dataset, which collects the time-dependent velocity profile of a pulsating Poiseuille Flow. As described in the \href{https://www.youtube.com/watch?v=8fhupzhAR_M}{\textcolor{blue}{second video tutorial}}, this dataset can be analytically decomposed in eigenfunction and hence offers a comparison between data-driven and analytical decompositions. Moreover, being the flow sustained by two known source terms, the exercise allows for exploring the convergence and the time-frequency analysis capabilities of all the methods. The second and third exercises present the analysis of 2D scalar datasets. These were described in \cite{MendezJournal2}. Both are useful to analyze the problem of uniqueness of the POD, which occurs when modes have similar energy content. The second exercise consists of a simple superposition of known modes while the third features the numerical solution of the nonlinear advection-diffusion of prescribed source terms. The fourth exercise, also presented in \cite{MendezJournal2}, presents the decomposition of the experimental data, which is velocity fields obtained via Time-Resolved Particle Image Velocimetry (TR-PIV). This allows for practice with modern experimental data. Figure \ref{fig:modes_jet} shows an exemplary mPOD mode obtained in MODULO for this test case, which consists of a planar gas jet impinging on a flat wall. The top figure shows the spatial structure of the mode; the bottom one shows the frequency content of the associated temporal structure. This mode captures the turbulent structures evolving from the shear layer instability in the jet.

As described in the video tutorials \href{https://www.youtube.com/watch?v=eG17tRQxqhk}{\textcolor{blue}{three}} and \href{https://www.youtube.com/watch?v=srcU_jfalXY}{\textcolor{blue}{seven}}, neither the POD nor the DFT can clearly isolate these structures. The POD is limited by the constraint of optimal convergence, which forces the decomposition to put multiple features in the same modes. The second is limited by the constraint of harmonic temporal structures, which does not let the DFT mode capture coherent patterns composed of multiple frequencies. Finally, the fifth exercise considers the velocity field obtained via TR-PIV of the flow past a cylinder in transient conditions. The dataset is described in \cite{DANTEC_VKI}. This test case consists of a much larger number of snapshots, which forces most laptop computers to use the memory saving features of MODULO.


\section{Impact and Conclusions}\label{Concl}


We have presented the functionalities of the open source software package MODULO, starting from the theoretical background. The software allows for performing classical modal decomposition such as POD and DFT as well as the novel mPOD. Moreover, thanks to its memory saving feature, MODULO is well suited to analyze relatively large data sets while keeping moderate memory requirements. While these decomposition are nowadays essential tools in fluid mechanics, their general framework is certainly of great interest to any applied scientist. Finally, the complete set of exercises available can also serve didactic purposes and encourage the novice to enter this important discipline.





\section*{References}
\bibliographystyle{elsarticle-num}
\bibliography{SoftwareX_Sub2_Ninni_Mendez}





\clearpage


\section*{Required Metadata}
\label{}

\section*{Current code version}
\label{}


\begin{table}[!h]
\begin{tabular}{|l|p{6.5cm}|p{6.5cm}|}
\hline
\textbf{Nr.} & \textbf{Code metadata description} & \textbf{} \\
\hline
C1 & Current code version & v1.1.1 \\
\hline
C2 & Permanent link to code/repository used for this code version & github.com/mendezVKI/MODULO \\
\hline
C3 & Legal Code License   & GNU General Public License v3.0 \\
\hline
C4 & Code versioning system used & None \\
\hline
C5 & Software code languages, tools, and services used & Matlab and Python \\
\hline
C6 & Compilation requirements, operating environments \& dependencies & 

Running the source code requires \textsc{Matlab 2017b} or higher.
Required Toolboxes:
Statistics and Machine Learning Toolbox v11.6, Signal Processing Toolbox v8.3 and Image Processing Toolbox v11.0).

The executable can be installed by any Microsoft Windows user.
\\
\hline
C7 & If available Link to developer documentation/manual & github.com/mendezVKI/MODULO \\
\hline
C8 & Support email for questions & davide.ninni@poliba.it, mendez@vki.ac.be \\
\hline
\end{tabular}
\caption{Code metadata}
\label{} 
\end{table}

\section*{Current executable software version}
\label{}


\begin{table}[!h]
\begin{tabular}{|l|p{6.5cm}|p{6.5cm}|}
\hline
\textbf{Nr.} & \textbf{(Executable) software metadata description} & \textbf{} \\
\hline
S1 & Current software version & v1.1.1 \\
\hline
S2 & Permanent link to executables of this version  & github.com/mendezVKI/MODULO/releases \\
\hline
S3 & Legal Software License & GNU General Public License v3.0 \\
\hline
S4 & Computing platforms/Operating Systems & Microsoft Windows \\
\hline
S5 & Installation requirements \& dependencies & None \\
\hline
S6 & If available, link to user manual - if formally published include a reference to the publication in the reference list & github.com/mendezVKI/MODULO \\
\hline
S7 & Support email for questions & davide.ninni@poliba.it, mendez@vki.ac.be \\
\hline
\end{tabular}
\caption{Software metadata}
\label{} 
\end{table}

\end{document}